
\input epsf.tex
%
\headline={\ifnum\pageno=1\firstheadline\else
\ifodd\pageno\rightheadline \else\leftheadline\fi\fi}
\def\firstheadline{\hfil}
\def\rightheadline{\hfil}
\def\leftheadline{\hfil}
        \footline={\ifnum\pageno=1\firstfootline\else\otherfootline\fi}
\def\firstfootline{\rm\hss\folio\hss}
\def\otherfootline{\hfil}

\font\tenrm=cmr10
\font\tenit=cmti10
\font\elevenbf=cmbx10 scaled\magstep 1
\font\elevenrm=cmr10 scaled\magstep 1
\font\elevenit=cmti10 scaled\magstep 1

\def\pmb#1{\setbox0=\hbox{#1}%
  \hbox{\kern-.025em\copy0\kern-\wd0
  \kern.05em\copy0\kern-\wd0
  \kern-0.025em\raise.0433em\box0} }

\catcode`@=11
\def\leftrightarrowfill{$\m@th\mathord\leftarrow \mkern-6mu
  \cleaders\hbox{$\mkern-2mu \mathord- \mkern-2mu$}\hfill
  \mkern-6mu \mathord\rightarrow$}
\def\overleftrightarrow#1{\vbox{\ialign{##\crcr
     \leftrightarrowfill\crcr\noalign{\kern-1pt\nointerlineskip}
     $\hfil\displaystyle{#1}\hfil$\crcr}}}
\catcode`@=12

\def\approxge{\hbox {\hfil\raise .4ex\hbox{$>$}\kern-.75 em
\lower .7ex\hbox{$\sim$}\hfil}}
\def\approxle{\hbox {\hfil\raise .4ex\hbox{$<$}\kern-.75 em
\lower .7ex\hbox{$\sim$}\hfil}}

\def \abstract#1 {\vskip 0.5truecm\sepline\vskip 0.5truecm
$$\vbox{\hsize=15truecm\noindent #1}$$}
\def \SISSA#1#2 {\vfil\vfil\centerline{Ref. S.I.S.S.A. #1 CM (#2)}}
\def \PACS#1 {\vfil\line{\hfil\hbox to 15truecm{PACS numbers: #1 \hfil}\hfil}}

\def \hfigure
     #1#2#3       {\midinsert \vskip #2 truecm $$\vbox{\hsize=14.5truecm
             \seven\baselineskip=10pt\noindent {\bcp \noindent Figure  #1}.
                   #3 } $$ \vskip -20pt \endinsert }

\def \hfiglin
     #1#2#3       {\midinsert \vskip #2 truecm $$\vbox{\hsize=14.5truecm
              \seven\baselineskip=10pt\noindent {\bcp \hfil\noindent
                   Figure  #1}. #3 \hfil} $$ \vskip -20pt \endinsert }

\def \vfigure
     #1#2#3#4     {\dimen0=\hsize \advance\dimen0 by -#3truecm
                   \midinsert \vbox to #2truecm{ \seven
                   \parshape=1 #3truecm \dimen0 \baselineskip=10pt \vfill
                   \noindent{\bcp Figure #1} \pretolerance=6500#4 \vfill }
                   \endinsert }

%
\def \ref
     #1#2         {\smallskip \item{[#1]}#2}
\def \sepline     {\medskip\centerline{\vbox{\hrule width5truecm}} \medskip}

\def \tabrul2     {\noalign{\vskip 5truept \hrule \vskip 2truept \hrule
                   \vskip 5truept} }


\footline={\ifnum\pageno>0 \tenrm \hss \folio \hss \fi }

\def\today
 {\count10=\year\advance\count10 by -1900 \number\day--\ifcase
  \month \or Jan\or Feb\or Mar\or Apr\or May\or Jun\or
             Jul\or Aug\or Sep\or Oct\or Nov\or Dec\fi--\number\count10}

\def\hour{\count10=\time\count11=\count10
\divide\count10 by 60 \count12=\count10
\multiply\count12 by 60 \advance\count11 by -\count12\count12=0
\number\count10 :\ifnum\count11 < 10 \number\count12\fi\number\count11}

\def\draft{
   \baselineskip=20pt
   \def\makeheadline{\vbox to 10pt{\vskip-22.5pt
   \line{\vbox to 8.5pt{}\the\headline}\vss}\nointerlineskip}
   \headline={\hfill \seven {\bcp Draft version}: today is \today\ at \hour
              \hfill}
          }

\catcode`@=11
%
%
\def\b@lank{ }

\newif\if@simboli
\newif\if@riferimenti

\newwrite\file@simboli
\def\simboli{
    \immediate\write16{ !!! Genera il file \jobname.SMB }
    \@simbolitrue\immediate\openout\file@simboli=\jobname.smb}

\newwrite\file@ausiliario
\def\riferimentifuturi{
    \immediate\write16{ !!! Genera il file \jobname.AUX }
    \@riferimentitrue\openin1 \jobname.aux
    \ifeof1\relax\else\closein1\relax\input\jobname.aux\fi
    \immediate\openout\file@ausiliario=\jobname.aux}

\newcount\eq@num\global\eq@num=0
\newcount\sect@num\global\sect@num=0

\newif\if@ndoppia
\def\numerazionedoppia{\@ndoppiatrue\gdef\la@sezionecorrente{\the\sect@num}}

\def\se@indefinito#1{\expandafter\ifx\csname#1\endcsname\relax}
\def\spo@glia#1>{} 

\newif\if@primasezione
\@primasezionetrue

\def\s@ection#1\par{\immediate
    \write16{#1}\if@primasezione\global\@primasezionefalse\else\goodbreak
    \vskip\spaziosoprasez\fi\noindent
    {\bf#1}\nobreak\vskip\spaziosottosez\nobreak\noindent}
%

\def\sezpreset#1{\global\sect@num=#1
    \immediate\write16{ !!! sez-preset = #1 }   }

\def\spaziosoprasez{50pt plus 60pt}
\def\spaziosottosez{15pt}

\def\sref#1{\se@indefinito{@s@#1}\immediate\write16{ ??? \string\sref{#1}
    non definita !!!}
    \expandafter\xdef\csname @s@#1\endcsname{??}\fi\csname @s@#1\endcsname}


\def\adv#1{\global\advance\sect@num by #1}


\def\autosez#1#2\par{
    \global\advance\sect@num by 1\if@ndoppia\global\eq@num=0\fi
    \xdef\la@sezionecorrente{\the\sect@num}
    \def\usa@getta{1}\se@indefinito{@s@#1}\def\usa@getta{2}\fi
    \expandafter\ifx\csname @s@#1\endcsname\la@sezionecorrente\def
    \usa@getta{2}\fi
    \ifodd\usa@getta\immediate\write16
      { ??? possibili riferimenti errati a \string\sref{#1} !!!}\fi
    \expandafter\xdef\csname @s@#1\endcsname{\la@sezionecorrente}
    \immediate\write16{\la@sezionecorrente. #2}
    \if@simboli
      \immediate\write\file@simboli{ }\immediate\write\file@simboli{ }
      \immediate\write\file@simboli{  Sezione
                                  \la@sezionecorrente :   sref.   #1}
      \immediate\write\file@simboli{ } \fi
    \if@riferimenti
      \immediate\write\file@ausiliario{\string\expandafter\string\edef
      \string\csname\b@lank @s@#1\string\endcsname{\la@sezionecorrente}}\fi
    \goodbreak\vskip 48pt plus 60pt
\centerline{\lltitle #2}                     
\par\nobreak\vskip 15pt \nobreak\noindent}

\def\semiautosez#1#2\par{
    \gdef\la@sezionecorrente{#1}\if@ndoppia\global\eq@num=0\fi
    \if@simboli
      \immediate\write\file@simboli{ }\immediate\write\file@simboli{ }
      \immediate\write\file@simboli{  Sezione ** : sref.
          \expandafter\spo@glia\meaning\la@sezionecorrente}
      \immediate\write\file@simboli{ }\fi
\noindent\lltitle \s@ection#2 \par}


\def\eqpreset#1{\global\eq@num=#1
     \immediate\write16{ !!! eq-preset = #1 }     }

\def\eqref#1{\se@indefinito{@eq@#1}
    \immediate\write16{ ??? \string\eqref{#1} non definita !!!}
    \expandafter\xdef\csname @eq@#1\endcsname{??}
    \fi\csname @eq@#1\endcsname}

\def\eqlabel#1{\global\advance\eq@num by 1
    \if@ndoppia\xdef\il@numero{\la@sezionecorrente.\the\eq@num}
       \else\xdef\il@numero{\the\eq@num}\fi
    \def\usa@getta{1}\se@indefinito{@eq@#1}\def\usa@getta{2}\fi
    \expandafter\ifx\csname @eq@#1\endcsname\il@numero\def\usa@getta{2}\fi
    \ifodd\usa@getta\immediate\write16
       { ??? possibili riferimenti errati a \string\eqref{#1} !!!}\fi
    \expandafter\xdef\csname @eq@#1\endcsname{\il@numero}
    \if@ndoppia
       \def\usa@getta{\expandafter\spo@glia\meaning
       \la@sezionecorrente.\the\eq@num}
       \else\def\usa@getta{\the\eq@num}\fi
    \if@simboli
       \immediate\write\file@simboli{  Equazione
            \usa@getta :  eqref.   #1}\fi
    \if@riferimenti
       \immediate\write\file@ausiliario{\string\expandafter\string\edef
       \string\csname\b@lank @eq@#1\string\endcsname{\usa@getta}}\fi}

\def\autoeqno#1{\eqlabel{#1}\eqno(\csname @eq@#1\endcsname)}
\def\autoleqno#1{\eqlabel{#1}\leqno(\csname @eq@#1\endcsname)}
\def\eqrefp#1{(\eqref{#1})}


\def\eq{\autoeqno}
\def\req{\eqrefp}



\newcount\cit@num\global\cit@num=0

\newwrite\file@bibliografia
\newif\if@bibliografia
\@bibliografiafalse

\def\lp@cite{[}
\def\rp@cite{]}
\def\trap@cite#1{\lp@cite #1\rp@cite}
\def\lp@bibl{[}
\def\rp@bibl{]}
\def\trap@bibl#1{\lp@bibl #1\rp@bibl}

\def\refe@renza#1{\if@bibliografia\immediate        
    \write\file@bibliografia{
    \string\item{\trap@bibl{\cref{#1}}}\string
    \bibl@ref{#1}\string\bibl@skip}\fi}

\def\ref@ridefinita#1{\if@bibliografia\immediate\write\file@bibliografia{
    \string\item{?? \trap@bibl{\cref{#1}}} ??? tentativo di ridefinire la
      citazione #1 !!! \string\bibl@skip}\fi}

\def\bibl@ref#1{\se@indefinito{@ref@#1}\immediate
    \write16{ ??? biblitem #1 indefinito !!!}\expandafter\xdef
    \csname @ref@#1\endcsname{ ??}\fi\csname @ref@#1\endcsname}

\def\c@label#1{\global\advance\cit@num by 1\xdef            
   \la@citazione{\the\cit@num}\expandafter
   \xdef\csname @c@#1\endcsname{\la@citazione}}

\def\bibl@skip{\vskip +4truept}


\def\stileincite#1#2{\global\def\lp@cite{#1}\global   
    \def\rp@cite{#2}}                                 
\def\stileinbibl#1#2{\global\def\lp@bibl{#1}\global   
    \def\rp@bibl{#2}}                                 

\def\citpreset#1{\global\cit@num=#1
    \immediate\write16{ !!! cit-preset = #1 }    }

\def\autobibliografia{\global\@bibliografiatrue\immediate
    \write16{ !!! Genera il file \jobname.BIB}\immediate
    \openout\file@bibliografia=\jobname.bib}

\def\cref#1{\se@indefinito                  
   {@c@#1}\c@label{#1}\refe@renza{#1}\fi\csname @c@#1\endcsname}

\def\cite#1{\trap@cite{\cref{#1}}}                  
\def\ccite#1#2{\trap@cite{\cref{#1},\cref{#2}}}     
\def\ncite#1#2{\trap@cite{\cref{#1}--\cref{#2}}}    
\def\upcite#1{$^{\,\trap@cite{\cref{#1}}}$}               
\def\upccite#1#2{$^{\,\trap@cite{\cref{#1},\cref{#2}}}$}  
\def\upncite#1#2{$^{\,\trap@cite{\cref{#1}-\cref{#2}}}$}  

\def\clabel#1{\se@indefinito{@c@#1}\c@label           
    {#1}\refe@renza{#1}\else\c@label{#1}\ref@ridefinita{#1}\fi}

\def\biblskip#1{\def\bibl@skip{\vskip #1}}           

\def\insertbibliografia{\if@bibliografia             
    \immediate\write\file@bibliografia{ }
    \immediate\closeout\file@bibliografia
    \catcode`@=11\input\jobname.bib\catcode`@=12\fi}


\def\commento#1{\relax}
\def\biblitem#1#2\par{\expandafter\xdef\csname @ref@#1\endcsname{#2}}


\catcode`@=12

\tolerance 100000
\biblskip{+8truept}                        
\def\hbup{\hfill\break\baselineskip 14pt}  


\global\newcount\notenumber \global\notenumber=0
\def\note #1 {\global\advance\notenumber by1 \baselineskip 10pt
              \footnote{$^{\the\notenumber}$}{\nine #1} \interlinea}



\font\scal=cmsy5
 \font\ltitle=cmbx10 scaled \magstep2



\def\gtrsim{\ \rlap{\raise 2pt \hbox{$>$}}{\lower 2pt \hbox{$\sim$}}\ }
\def\lesssim{\ \rlap{\raise 2pt \hbox{$<$}}{\lower 2pt \hbox{$\sim$}}\ }


\def\scss{\scriptscriptstyle}



\def\ea{{\elevenit et.al.}}
\def\ib{{\elevenit ibid.\ }}

\def\npb#1{{\elevenit Nucl. Phys.} {\elevenbf B#1},}
\def\plb#1{{\elevenit Phys. Lett.} {\elevenbf B#1},}
\def\prd#1{{\elevenit Phys. Rev.} {\elevenbf D#1},}
\def\prl#1{{\elevenit Phys. Rev. Lett.} {\elevenbf #1},}

\def\zpc#1{{\elevenit Z. Phys.} {\elevenbf C#1},}
\def\prep#1{{\elevenit Phys. Rep.} {\elevenbf #1},}


\stileincite{}{}     
\stileinbibl{}{.}    

\numerazionedoppia   



\def\scss#1{{\scriptscriptstyle #1}}


\def\slesq{\left(s_L^e\right)^2}
\def\slnuesq{\left(s_L^{\nu_e}\right)^2}
\def\slmusq{\left(s_L^\mu\right)^2}
\def\slnumusq{\left(s_L^{\nu_\mu}\right)^2}
\def\sresq{\left(s_R^e\right)^2}

\def\srmusq{\left(s_R^\mu\right)^2}

\def\slusq{\left(s_L^u\right)^2}
\def\sldsq{\left(s_L^d\right)^2}
\def\slssq{\left(s_L^s\right)^2}
\def\slcsq{\left(s_L^c\right)^2}

\def\sltausq{\left(s_L^\tau\right)^2}
\def\srtausq{\left(s_R^\tau\right)^2}
\def\srusq{\left(s_R^u\right)^2}
\def\srdsq{\left(s_R^d\right)^2}

\def\srcsq{\left(s_R^c\right)^2}
\def\slbsq{\left(s_L^b\right)^2}

\def\slnutausq{\left(s_L^{\nu_\tau}\right)^2}


\def\G{{\cal G_{\rm SM}}}

\def\N{{\hbox{\scal N}}}
\def\K{{\hbox{\scal K}}}


\def\pr{\prime}




\def\gweak{SU(2)_L\times U(1)_Y}
\autobibliografia
\pageno=0
\line{\hfil }
\vglue 1cm
\hsize=6.0truein
\vsize=8.5truein
\parindent=3pc
\baselineskip=10pt
\null
\rightline{UM-TH 93--14}\qquad\par\noindent
\rightline{hep-ph@xxx/yymmddd}\par\noindent
\vglue 1.0cm
\centerline
{\ltitle REVIEW OF CONSTRAINTS ON FERMION MIXING$^{\displaystyle\dagger}$ }
\vglue 1.5cm
\centerline{\elevenrm ENRICO NARDI }
\baselineskip=14pt
\vglue 0.3cm
\centerline{\elevenit Randall Laboratory of Physics, University of Michigan}
\smallskip
\baselineskip=13pt
\centerline{\elevenit Ann Arbor, MI 48109-1120, U.S.A.}
\vglue 2.0cm
\centerline{\elevenrm ABSTRACT}
\vglue 0.6cm
{\rightskip=1pc
 \leftskip=1pc
 \elevenrm\baselineskip=14pt
 \noindent
The constraints on the mixing angles of the standard fermions with new
heavy particles that can appear in many extensions of the electroweak
theory are reviewed. Some emphasis is put in distinguishing the
effects of a mixing with new states which transform in a non-canonical
way with respect to $SU(2)\times U(1)$ (left-handed singlets or
right-handed doublets), from the effects of a mixing with new states
with standard quantum number assignments. Constraints from flavor
changing neutral current processes, as well as from flavor diagonal
and charged current experimental data are considered. New limits for
lepton flavor violating mixings are presented.
By using the most
recent results on the $\tau$ mass, lifetime and branching ratios,
updated limits on the
mixing parameters of the $\tau$ neutrino are derived.
These limits are improved up to a factor of 3, and no signals
of deviation from the standard model predictions are found.
Limits on the mass of a $Z_1$ gauge boson from E$_6$, from the absence
of muon number violating processes are briefly discussed as well.
\vglue 0.6cm}
\noindent
PACS number(s): 13.10.+q,12.10.Dm,12.15.Ff,14.60.Jj
\vfill
\noindent
--------------------------------------------\phantom{-} \hfil\break
\leftline{$^\dagger$ Invited talk at the
``Workshop on Physics and Experiments at Linear e+e- Colliders'',}
\leftline{\phantom{$^\dagger$} Waikoloa, Hawai, April 26-30, 1993.}
\vglue 0.2cm
\leftline{E-mail: nardi@umiphys.bitnet}
\bigskip
\leftline{UM-TH 93--14}
                   \bigskip
\centerline{July 1993}

\bigskip \bigskip
\eject


\phantom{\centerline{\elevenbf REVIEW OF CONSTRAINTS ON FERMION MIXING
}}
\vglue 1.0cm
\phantom{\centerline{\tenrm ENRICO NARDI }}
\baselineskip=13pt
\phantom{
\centerline{\tenit Randall Laboratory of Physics, University of
Michigan}}
\baselineskip=12pt
\phantom{
\centerline{\tenit Ann Arbor, MI 48109-1120, U.S.A.}}
\vglue 0.8cm
\adv{1}
\line{\elevenbf 1. Introduction \hfil}
\bigskip
\baselineskip=14pt
\elevenrm
\noindent
The Standard Model (SM) of the electroweak interactions has achieved a
tremendous success in describing the experimental data within the
range of energies available today,
however too many questions are still left unanswered, and it is
generally believed that the present theory cannot be the whole
story.
Several models which go beyond the SM and
 which do address some of these remaining questions
predict the existence of new particles. In particular, in
models based on gauge groups larger than the SM group
$\G = SU(2)_L \times U(1)_Y \times SU(3)_C$
the new gauge interactions require the presence of new fermions to
insure anomaly cancellation.

In the present analysis I will be only interested
in fermions with conventional electric charges, implying that
the new states must be singlets or doublets of weak--isospin.
A rather heterodox exception is that of a gauge triplet of
fermions\upcite{trip}, but this
possibility will not be considered here.
The possibilities for the {\elevenit new} fermions are listed in Table 1.
Vector singlet
(doublet) fermions refer to particles whose $L$ and $R$ components both
transform as singlets (doublets) under $SU(2)_L$.
$\gweak$ singlet Weyl neutrinos can also be present.
A typical example in which these new states
appear is represented by E$_6$ models, that will be briefly discussed in
Section 3.
Mirror fermions are another
type of new fermion, whose transformation properties under $SU(2)_L$ are
opposite those of ordinary fermions, i.e.~L-handed singlets and
R-handed doublets. They appear, for instance, in grand unified
theories which include family unification\upcite{mirror}.
\midinsert
{
$$
\vbox{\hsize= 13.35truecm
{\tenrm \baselineskip=12pt
\noindent
Table 1:
Possible $\gweak$ assignments for new fermions. Pairs
of particles enclosed in parentheses indicate $SU(2)_L$-doublets, otherwise
they are $SU(2)_L$-singlets. $N$ and $E$ refer to leptons of charge $0$ and
$-1$, respectively. $U$ and $D$ are quarks of charge $2/3$ and $-1/3$.}
\vskip -.2truecm}
$$
$$
\vbox{\offinterlineskip
\halign{&\vrule#&
   \strut\quad#\hfil\quad\cr
\noalign{\hrule}
height2pt&\omit&\cr
&~~~~~{\elevenit a) Vector Doublets}&\cr
&~~~~~~~${N\choose E^-}_L~~~{N\choose E^-}_R~,
      ~~~{U\choose D}_L~~~{U\choose D}_R$&\cr
height2pt&\omit&\cr
&~~~~~{\elevenit b) Vector Singlets}&\cr
&~~~~~~~$E^-_L~~~E^-_R~,~~~U_L~~~U_R~~~D_L~~~D_R$&\cr
height2pt&\omit&\cr
&~~~~~{\elevenit c) Weyl Neutrinos}&\cr
&~~~~~~~$N_L~~~N_R$&\cr
height2pt&\omit&\cr
&~~~~~{\elevenit d) Mirror Fermions}&\cr
&~~~~~~~$E_L^-~~~{N\choose E^-}_R~,~~~{U_L\atop D_L}~~~{U\choose D}_R$&\cr
height2pt&\omit&\cr
\noalign{\hrule}}}
$$
}
\endinsert
There are two ways to search for signals of new fermions:
directly and indirectly.
Model independent limits from direct production come
from LEP\upcite{pdg92} and set a lower bound on the masses of such
particles of $m_{new}\gtrsim 45~{\rm GeV}\,$,
although
the mass limit on singlet neutrinos can be considerably weaker. As
to indirect signals, one possibility is to look for loop-induced effects in
high precision measurements or in
rare processes. This are model-dependent analyses, depending on the
number, masses and couplings
of the new states, and will not be discussed here.
The other possibility, which is the topic of this brief review,
is to search for the new states by looking
for signals of mixings of the new fermions
with the known ones. This possibility
relies on the fact that the most general
gauge invariant Lagrangian involving the new particles usually
leads to mass eigenstates which correspond to some
superposition of the known and new gauge eigenstates.
These mixings can be analysed in a model independent way.

In Section 2 I will briefly outline the formalism for dealing with
mixing effects for the charged fermions in the
neutral current (NC) sector.
For a more complete treatement including the
charged currents (CC) and
the mixings of the  neutral states I refer to the
original works  by Langacker and London\upcite{ll1}, and
Nardi, Roulet and Tommasini\upccite{fit}{fit6}.
In Section 3 I will review the
constraints which current experimental data place upon the mixings
between fermions with different $SU(2)_L$ transformation properties,
and I will also describe the kind of constraints that,
in the frame of E$_6$ models,  can be
set on the $Z_1$  parameters from an analysis of some
unsuppressed flavor changing (FC) mixings of the known leptons.
\vglue 0.6cm
\adv{1}
\line{\elevenbf 2. Formalism \hfil}    
\vglue 0.4cm
\noindent
I will assume an effective low energy gauge group of the form
$\G\times U_1(1)$
where, if the rank of the group is larger than 5, as in E$_6$,
the additional abelian factor
arises as a combination of different Cartan generators, and
corresponds to the lightest additional neutral gauge boson.
Then the neutral current Lagrangian in the gauge basis reads
$$
-{\cal L}_{\rm NC}=eJ^\mu_{\rm em}A_\mu +
g_0 J_0^\mu Z_{0 \mu} + g_1 J_1^\mu Z_{1 \mu}.
\eq{2.1}
$$
The SM neutral gauge boson $Z_0$ couples with strength
$g_0=(4\sqrt{2} G_F M^2_{Z_0})^{1/2}$ to the usual combination of the
neutral isospin and electromagnetic currents
$
J^\mu_0=J^\mu_3-\sin^2\theta_W  J^\mu_{\rm em}.
$
Assuming that the new $U_1(1)$ originates from a GUT based on a
{\elevenit simple} group, and normalizing  the new generator $Q_1$ to the
hypercharge axis, the $Z_1$ couples to the new $J_1$ current with
strength
$g_1 \simeq g_0 \sin\theta_W$.
In general the standard  $Z_0$ is expected to be
mixed with the $Z_1$, however in this short presentation I will
neglect these additional
effects, since due to the  tight limits implied
for the $Z_0$--$Z_1$ mixing by low energy NC and
LEP data ($\phi_{Z_0-Z_1}\lesssim 0.02\>$\upccite{fit6}{zp-new})
they turn out to be less important than the effects due to direct
$Z_1$ exchange\upcite{lfc}.
To ensure the absence of
anomalies for the new gauge current $J_1\,$, {\elevenit new} fermions
must be present in addition to the standard 15 {\elevenit known}
fermions per generation. Here I will assume that some of the
additional new fermions are electrically charged, and that they are
mixed with the known states. Each of the conventional {\elevenit
light} fermion mass eigenstate then corresponds to a superposition of
the known states and the new states. Conservation of the electric and color
charges forbids a mixing between gauge eigenstates with different
$U(1)_{\rm em}$ and $SU(3)_{\rm c}$ quantum numbers, implying in turn
that the corresponding currents are not modified by the presence of
the new states. In contrast the neutral isospin generator $T_3$ and
the new generator $Q_1$ are spontaneously broken, and a mixing between
states with different $t_3$ and $q_1$ eigenvalues is allowed. This
will affect the $J_3$ and $J_1$ currents\upcite{fit6}
and in turn the couplings of
the light mass eigenstates to the $Z_0$ and $Z_1$.
In the gauge currents chirality is conserved too, and it is then
convenient to group the fermions with the same electric charge and
chirality $\alpha=L,R$ in a vector of the known ($\cal K$) and new
($\cal N$) gauge eigenstates  $\Psi^{o}_\alpha=(\Psi^o_{\K},
\Psi^o_{\N})_\alpha^T$. This vector is related to the corresponding vector
of the light ({\elevenit l}$\,$) and heavy ({\elevenit h}) mass eigenstates
$\Psi_\alpha=(\Psi_l,\Psi_h)_\alpha^T$ through a unitary transformation
$$
\pmatrix{\Psi^o_{\K}\cr\Psi^o_{\N}}_\alpha = U_\alpha
\pmatrix{\Psi_l\cr\Psi_h}_\alpha \qquad{\rm where}\qquad
U_\alpha = \pmatrix{A &G\cr F & H}_\alpha ,
\qquad  \alpha=L,R.
\eq{2.2}
$$
The submatrices $A$ and $F$ describe the overlap of the light
eigenstates with the known and the new states respectively, and
the unitarity of $U_\alpha$ implies
$
A^\dagger A+F^\dagger F=A A^\dagger +G G^\dagger =I.
$
Note that there is no extra index to
label the electric charge, nevertheless
$\Psi^o_\alpha$ and $\Psi_\alpha$
will  be treated
as vectors corresponding to a definite value of $q_{\rm em}$.
In terms of the fermion mass eigenstates, the neutral current
corresponding to a (broken) generator ${\cal Q}=T_3,Q_1$ now reads
$$
J^\mu_{\cal Q} =\sum_{\alpha=L,R}
\bar\Psi_{\alpha} \gamma^\mu U^\dagger_\alpha {\cal Q}_\alpha
U_\alpha\Psi_{\alpha},
\eq{2.4}
$$
where ${\cal Q}_\alpha$ represents a generic diagonal matrix of the
charges $q_\alpha=t_3(f_\alpha)$, $q_1(f_\alpha)$ for the chiral
fermion $f_\alpha$. Since we are interested in the indirect effects of
fermion mixings in the couplings of the light mass eigenstates, we
have to project $J^\mu_{\cal Q}$ onto the light components $\Psi_l$.
In the particularly simple case when the mixing is with only
one type of new fermions with the same $q_\alpha^\N$ charges,
by means of the unitarity relations for $U_\alpha$
we easily obtain\upcite{fit6}
$$
J^\mu_{l{\cal Q}}
               =\sum_{\alpha=L,R}
\bar \Psi_{l\alpha} \gamma^\mu \left[ q_\alpha^{\K} I +
(q_\alpha^{\N} - q_\alpha^{\K})
F^\dagger_\alpha F_\alpha \right]\Psi_{l\alpha}.
\eq{2.5}
$$
In Eq. \req{2.5}
$q_\alpha^{\K}I $ represents the coupling of a particular light
fermion in the absence of mixing effects, while the second term
accounts for the modifications due to fermion mixings. The matrix
$F_\alpha^\dagger F_\alpha$ is in general not diagonal, and clearly
whenever the coefficient $(q_\alpha^{\N} - q_\alpha^{\K})$ is nonvanishing,
the off diagonal terms will induce FCNC, while the diagonal terms
will affect the flavor-conserving couplings.
It is useful to parametrize the FC mixing between the light particles $i$ and
$j$ as
$$
{\cal F}_\alpha^{ij} = \left(F_\alpha^\dagger F_\alpha\right)_{ij},
\hskip 1.5truecm   i\ne j,  \qquad  \qquad  \alpha=L,R.
\eq{2.6}
$$
{}From a phenomenological
point of view
we can now distinguish two cases. If the
known states are mixed with
new particles whose $L$ and/or $R$ components
transform in a noncanonical way under $SU(2)_L$
(and hence will be referred to as {\elevenit exotic})
weak--isospin is violated:
$t_3(f_\alpha^{\N}) \ne t_3(f_\alpha^{\K})$. Then the $J_0$ current
is affected and the $Z_0$ interactions will be FC.
If the mixing is with new states which
obey the same transformation
properties than the known  fermions
(i.e.~$L$-handed doublets and $R$-handed singlets, that will be referred
to as  {\elevenit ordinary}) then
$t_3(f_\alpha^{\N}) =
t_3(f_\alpha^{\K})$, and clearly  the $J_0$ current is not
affected and in particular it remains flavor-diagonal.
However, in general we still
have $q_1(f_\alpha^{\N}) \ne q_1(f_\alpha^{\K})$. Then the
isospin-conserving mixings can indeed affect the $J_1$ current,
inducing sizeable FC couplings to the $Z_1$.

Now, to give an example of the possible form of the fermion mass matrices
and to discuss the expected magnitude of the mixings between the known and the
new fermions, let us introduce for each fermion family
a vector gauge singlet of new fermions
$({X^o_{\scss E}}_L,{X^o_{\scss O}}_R)_i$
($E$ = exotic, $O$ = ordinary, $i=1,2,3$)
with the same electric and color
charges than the  known fermions $({f^o_{\scss O}}_L,{f^o_{\scss
O}}_R)_i$. Then in the gauge eigenstate basis
the mass term reads
$$
{\cal L}_{\rm mass}=
{(\bar {f^o_{\scss O}}, \bar {X^o_{\scss E}})}_L \>
{\cal M} \> {f^o_{\scss O} \choose  X^o_{\scss O}}_R,
\hskip 1truecm
{\cal M} = \pmatrix{D &D^\pr\cr S^\pr &S},
\eq{2.7}
$$
where e.g. $f^o = (f^o_1,f^o_2,f^o_3)^T$ etc..
The entries
$D$ and $D^\pr$ in the non diagonal mass matrix ${\cal M}$, are
$3\times3$ matrices generated by vacuum expectation values (vevs) of
doublets multiplied by Yukawa couplings, while $S$ and $S^\pr$ are
generated by vevs of singlets.
As a general rule, while the mass terms which couple ordinary
L-fermions to ordinary R-fermions (or exotic L-fermions to exotic
R-fermions) arise from vevs of Higgs doublets, the entries which
couple ordinary fermions to the exotic ones are generated by vevs of
singlets.
Higgs singlets are responsible for the large masses of the
new heavy fermions in vector multiplets and, in most cases, also
contribute to the mass of the new heavy gauge boson; hence it is
natural to assume $S,S^\pr \gg D,D^\pr$.
In the presence of such a hierarchy among the entries in ${\cal M}$ it
is easy to see\upcite{lfc} that for the matrix $U_\alpha$ in
Eq. \req{2.2}
describing the
ordinary--exotic mixings between $L$-states
it is natural to expect that
the submatrices $F$ and $G$ would acquire an overall suppression
factor $\sim S/D \ll 1 $ of the order of the ratio of the light to
heavy mass scale. In contrast such a suppression is not present for
the ordinary-ordinary $F$ and $G$ mixing terms for the $R$-states\upcite{lfc}.
Now, since it is precisely $F^\dagger F$ in Eq. \req{2.5} which affects
the flavor diagonal couplings and also induces  FCNC,
if the mass scale for the exotic fermions is large enough, the suppression
of the ordinary--exotic mixings explains in a natural way the
non--observations of these effects in the $Z_0$ interactions.
On the other hand, for the ordinary--ordinary mixings there is no
reason to expect the elements of $F^\dagger F$ to be particularly
small\upcite{lfc} , and accordingly FC processes can be expected to
occur at a sizeable rate in $Z_1$ interactions.
Then, in the cases in which the FC couplings to the $Z_1$ are not
suppressed, the limits on FC processes point toward a rather heavy
$Z_1$\upccite{lfc}{mue}.

As can be seen
from Table 2, for $t_3(f_\alpha^{\N}) \ne t_3(f_\alpha^{\K})$
very stringent constraints can be derived for
${\cal F}_{L,R}^{ij},~(i\ne j)$
which strongly limits the mixing of ordinary and exotic
fermions. However, it is possible to evade these bounds by considering the
fine-tuned cases in which the mixing matrices
$F_\alpha^\dagger F_\alpha$ are
diagonal. These correspond to those directions in parameter space in
which each known ordinary fermion mixes with its own exotic fermion.
If the strong assumption ${\cal F}_{L,R}^{ij}=0$ for $i\ne j$
is made, one can write
$$
\left(F_\alpha^\dagger F_\alpha\right)_{ij}=
\left(s^i_\alpha\right)^2 \delta_ij \hskip 1truecm \alpha=L,R,
\eq{2.8}
$$
in which $\left(s^i_\alpha\right)^2 \equiv \sin^2\theta_\alpha^i$,
where $\theta_{L(R)}^i$ is the mixing angle of the
$i^{th}$ $L$-handed ($R$-handed) light fermion.
In this case the ordinary-exotic mixing is
parametrized by one angle per each $L$- or $R$-handed charged
fermion. The same angles enter also the expression for
the CC measurables.
We refer to Refs. [\cite{ll1},\cite{fit}] for the extension of the
formalism  to the CC sector.
Even if the limits on FCNC processes are not effective
to constrain these particular fine-tuned mixings,
still all the $\left(s^i_\alpha\right)^2$  can be constrained
by looking at the high precision data involving
CC and flavor-diagonal NC.
The corresponding limits are given in Table 3.

The formalism for the neutral sector is
more complicated, both due to the possibility of
Dirac and Majorana masses, and because there is no empirical evidence
requiring the absence of FCNC between neutrino species. However, due to the
fact that neutrinos are unobserved in experiments, it is possible to
parametrize their mixing by one effective angle for neutrino
flavor\upcite{ll1},
with the addition of an auxiliary
effective parameter $\Lambda$ that
takes into account the type of new neutrinos involved in the mixing.
If the mixing is purely with
new ordinary, exotic singlet or exotic doublet
neutrinos, we have respectively $\Lambda=$0,2,4.
Again we refer to Refs. [\cite{ll1},\cite{fit}] for
a detailed discussion of the mixings in the neutral sector.
\vglue 0.6cm
\adv{1}
\line{\elevenbf 3. Results \hfil}
\bigskip
\line{\elevenit 3.1.
Limits on flavor changing ordinary-exotic fermion mixing \hfil}
\smallskip\noindent
The limits on the
FCNC parameters
${\cal F}^{ij}$ Eq. \req{2.6}
which can be obtained assuming a mixing with
new exotic fermions are listed in Table 2.
The bounds on lepton flavor violating (LFV) NC
from the leptonic decays $\mu\not\to 3e$, $\tau\not\to 3e$, $3\mu$
are taken or adapted from Ref.[~\cite{lfc}].
The bound on ${\cal F}^{\mu e}$ from the non-observation of
$\mu$--$e$ conversion in nuclei is taken from
Ref.[~\cite{mue}].
The limits on $\vert{\cal F}^{\tau e}\vert$ and
$\vert{\cal F}^{\tau \mu}\vert$
from $Z_0\not\to {\tau e}$, ${\tau \mu}$
are new, and have been derived from the experimental results
given in the first paper in Ref. [\cite{lep-lfc}].
It is interesting to note that in all cases the best bounds
are obtained from processes different from leptonic decays.
The limits on hadronic FCNC
have been presented in Ref.~[\cite{london}] and
were taken, updated or adapted from
Ref.~[\cite{Silverman}].
There is no bound on
$\vert{\cal F}^{bd}\vert$ from $B_d^0$-${\overline{B_d^0}}$ mixing because
this mixing can in principle be explained by a non-zero
${\cal F}^{bd}$\upcite{Silverman}.
\midinsert
{
$$
\vbox{\hsize= 13.35truecm
{\tenrm \baselineskip=12pt
\noindent
Table 2:
Limits on the FCNC parameters
${\cal F}^{ij}$ for ordinary-exotic fermion mixing.
The bounds on leptonic FCNC
from leptonic decays are
taken or adapted from Ref.[~\cite{lfc}], the one from
$\mu N \not\to e N$ is taken from Ref.[~\cite{mue}],
and those from $Z_0\not\to \tau e$, $\tau \mu$ are new.
The limits on hadronic FCNC
have been presented in Ref.~[\cite{london}] and
were taken, updated or adapted from
Ref.~[\cite{Silverman}].
The experimental reference is given in the `Source' column.}
\vskip -.2truecm}
$$

$$
\vbox{\offinterlineskip
\halign{&\vrule#&
   \strut\quad#\hfil\quad\cr
\noalign{\hrule}
height2pt&\omit&&\omit&&\omit&\cr
& Quantity && Upper Limit && Source &\cr
height2pt&\omit&&\omit&&\omit& \cr
\noalign{\hrule}
height2pt&\omit&&\omit&&\omit&\cr
& $\vert{\cal F}^{\mu e}\vert$ && $2.4\times 10^{-6}$ &&
			$\mu\not\to 3e$ \quad[\cite{pdg92}] &\cr
& \omit && $1.0\times 10^{-5}$ &&
                        $\mu N \not\to e N$ \quad[\cite{mue-exp}] &\cr
& $\vert{\cal F}^{\tau e}\vert$ && $1.4\times 10^{-2}$ &&
			$\tau\not\to 3e$ \quad[\cite{pdg92}] &\cr
& \omit && $8.0\times 10^{-3}$ &&
                        $Z_0 \not\to \tau e $ \quad[\cite{lep-lfc}] &\cr
& $\vert{\cal F}^{\tau \mu}\vert$ && $1.1\times 10^{-2}$ &&
			$\tau\not\to 3\mu$ \quad[\cite{pdg92}] &\cr
& \omit && $9.5 \times 10^{-3}$ &&
                        $Z_0 \not\to \tau \mu $ \quad[\cite{lep-lfc}] &\cr
& $\vert{\cal F}^{ds}\vert$ && $6\times 10^{-4}$ &&
			$\Delta m_{K_L K_S}$ \quad[\cite{pdg92}] &\cr
& \omit && $1\times 10^{-5}$ && $K_L\to\mu^+\mu^-$ \quad[\cite{pdg92}] &\cr
& $\vert{\cal F}^{cu}\vert$ && $1\times 10^{-3}$ &&
		$D^0$-${\overline{D^0}}$ mixing \quad[\cite{pdg92}] &\cr
& $\vert{\cal F}^{bd}\vert$, $\vert{\cal F}^{bs}\vert$
     && $2\times 10^{-3}$ && $B\not\to\ell^+\ell^- X$ \quad[\cite{UA1-B}] &\cr
height2pt&\omit&&\omit&&\omit& \cr
\noalign{\hrule}}}
$$
}
\endinsert
\bigskip
\line{\elevenit 3.2.
Limits on flavor diagonal ordinary-exotic fermion mixing \hfil}
\smallskip\noindent
The limits on the flavor diagonal
NC parameters
$\left(s^i_{L,R}\right)^2$ Eq. \req{2.8} are listed in Table 3,
which is taken from Ref.~\cite{fit}.
They correspond to those fine-tuned direction in parameter space
for which the constraints from FCNC processes are evaded.
The results of two different fits are shown. In the
`individual fit' only one mixing angle at a time is allowed to be
non-zero, while in the `joint fit' all mixing angles
are simultaneously present.
In the `Source' column in Table 3 are listed those
observables which are most important for constraining the mixing angles in
the individual fits.
The $\Gamma$'s denote partial widths of the $Z_0$,
the $A$'s are asymmetries, $\nu e$ and $\nu q$ refer to low energy NC
scattering experiments, $e q$ refers to measurements of parity violation
in atoms, $g_\ell$ ($\ell=e$,$\mu$,$\tau$) are derived from lepton universality
in CC  processes, $V^2_{ui}$ refers to the unitarity of the CKM
matrix, while
$s_{eff}^{LEP}$ and $s_{eff}^{NC}$ refer to the weak mixing
angle as extracted in NC measurements at the $Z_0$ peak and at
low energy, respectively.
In the joint fit it is possible to evade the
bounds from these observables through fine-tuned cancellations between
different mixings,
so that the constraints in the joint fit are somewhat weaker than
those in the individual fit.
In this case other observables, which depend on
different combinations of the mixings
and which are denoted by a * in Table 3,
become important.
It should be stressed that the data used to obtain these constraints are
already a bit out of date. For example, only the 1990 LEP data was
used. The inclusion of the 1991-92 LEP data would surely strengthen
most of the bounds somewhat.
The most important new development is in $\tau$-decays.
The value of $(g_\tau/g_e)^2$ that was used in Ref. [\cite{fit}]
was about 1.5 standard deviations away from its SM value,
pointing towards a non-zero mixing for $\nu_\tau$.
In the present analysis the new values for the tau mass
$m_\tau=1776.9\pm 0.5\,$MeV\upcite{taumass},
for the $\tau$ lifetime $T_\tau=295.7\pm3.2\,$fs\upcite{taulife},
and for the $\tau$ leptonic branching fractions\upcite{taubranch}
$B(\tau\to e\bar\nu\nu)=(17.75\pm0.15)\%$ and
$B(\tau\to \mu\bar\nu\nu)=(17.39\pm0.15)\%$
have been used.
As a result any hint for a non-zero mixing of the $\tau$
neutrino has disappeared, while the limits on
$\left(s_L^{\nu_\tau}\right)^{2}$ have been improved up to a factor of 3.
\midinsert
{
$$
\vbox{\hsize= 6.0truein 
{\tenrm \baselineskip=12pt
\noindent
Table 3:
90\% C.L. upper limits on the ordinary-exotic flavor diagonal
mixing angles for individual fits (one
angle at a time is allowed to vary) and joint fits (all angles allowed to
vary simultaneously)\upcite{fit}. Observables which are most
important for the constraints are shown in the `Source' column.
The different values of the parameter $\Lambda$ correspond to
mixings with different kinds of new neutrinos, as explained in the
text. For the limits on $\slnutausq$
the most recent results on the $\tau$ mass\upcite{taumass},
lifetime\upcite{taulife} and branching
ratios\upcite{taubranch} have been used. }
\vskip -.2truecm}
$$

$$
\vbox{\offinterlineskip
\halign
{&\vrule#&
   \strut\quad#\hfil\quad\cr
\noalign{\hrule}
height2pt&\omit&&\omit&&\multispan5 &&\omit&\cr
& \omit && \omit\hidewidth Individual \hidewidth &&
				\multispan5 Joint && Source &\cr
height2pt&\omit&&\omit&&\multispan5 &&\omit&\cr
\noalign{\hrule}
height2pt&\omit&&\omit&&\omit&&\omit&&\omit&&\omit&\cr
& \omit && \omit && $\Lambda=2$ && $\Lambda=0$ && $\Lambda=4$ && \omit &\cr
height2pt&\omit&&\omit&&\omit&&\omit&&\omit&&\omit&\cr
\noalign{\hrule}
height2pt&\omit&&\omit&&\omit&&\omit&&\omit&&\omit&\cr
& $\slesq$ && 0.0047 && 0.015 && 0.0090 && 0.015 &&
			$\Gamma_e,M_W^*,A_\mu^{FB*},eq^*,g_e^*$ &\cr
& $\sresq$ && 0.0062 && 0.010 && 0.0082 && 0.010 &&
			$\Gamma_e,A_e^{FB},A_\mu^{FB*},\nu e^*$ &\cr
& $\slmusq$ && 0.0017 && 0.0094 && 0.0090 && 0.011 &&
		$V_{ui}^2,\nu q,g_\mu,\Gamma_\mu,s_{eff}^{LEP*}$ &\cr %
& $\srmusq$ && 0.0086 && 0.014 && 0.014 && 0.013 &&
			$\Gamma_\mu,A_\mu^{FB}$ &\cr
& $\sltausq$ && 0.011 && 0.017 && 0.015 && 0.017 &&
			$\Gamma_\tau,A_\tau^{FB},g_\tau,A_\tau^{pol*}$ &\cr
& $\srtausq$ && 0.011 && 0.012 && 0.014 && 0.012 &&
			$\Gamma_\tau,A_\tau^{pol},A_\tau^{FB},g_\tau^*$ &\cr
& $\slusq$ && 0.0045 && 0.019 && 0.015 && 0.019 &&
			$V_{ui}^2,\Gamma_h,\Gamma_Z,eq,\nu q$ &\cr
& $\srusq$ && 0.018 && 0.024 && 0.025 && 0.024 &&
			$\nu q,\Gamma_h,\Gamma_Z,eq$ &\cr
& $\sldsq$ && 0.0046 && 0.019 && 0.016 && 0.019 &&
			$V_{ui}^2,\Gamma_h,\Gamma_Z,\nu q$ &\cr
& $\srdsq$ && 0.020 && 0.030 && 0.028 && 0.029 &&
			$eq,\Gamma_h,\Gamma_Z,\nu q$ &\cr
& $\slssq$ && 0.011 && 0.038 && 0.039 && 0.041 &&
			$\Gamma_h,\Gamma_Z,V_{ui}^2$ &\cr
& $\left(s_R^s\right)^{2}$ && 0.36 && 0.67 && 0.63 && 0.74 &&
			$\Gamma_h,\Gamma_Z$ &\cr
& $\slcsq$ && 0.013 && 0.040 && 0.042 && 0.042 &&
			$\Gamma_h,\Gamma_Z,\Gamma_c^*,A_c^{\gamma Z*}$ &\cr
& $\srcsq$ && 0.029 && 0.097 && 0.10 && 0.099 &&
		$\Gamma_h,\Gamma_Z,A_c^{\gamma Z*},\Gamma_c^*,A_c^{FB*}$ &\cr
& $\slbsq$ && 0.011 && 0.070 && 0.072 && 0.069 &&
			$\Gamma_h,\Gamma_Z,\Gamma_b,A_b^{FB*}$ &\cr
& $\left(s_R^b\right)^{2}$ && 0.33 && 0.39 && 0.40 && 0.39 &&
		$\Gamma_b,\Gamma_Z,\Gamma_h,A_b^{\gamma Z},A_b^{FB*}$ &\cr
& $\slnuesq$ && 0.0097 && 0.015 && 0.016 && 0.014 &&
			$s_{eff}^{LEP},g_e,s_{eff}^{NC},M_W^*$ &\cr
& $\slnumusq$ && 0.0019 && 0.015 && 0.0087 && 0.011 &&
			$V_{ui}^2,g_\mu,\nu q,s_{eff}^{LEP},M_W^*$ &\cr
& $\left(s_L^{\nu_\tau}\right)^{2}$
	&& 0.023 && 0.033 && 0.034 && 0.026 && $\Gamma_Z,g_\tau$ &\cr
height2pt&\omit&&\omit&&\omit&&\omit&&\omit&&\omit&\cr
\noalign{\hrule}}}
$$
}
\endinsert
\line{\elevenit 3.3.  Limits on a Z$_1$ from E$_6$ from
ordinary-ordinary fermion mixing \hfil}
\smallskip\noindent
E$_6$ GUTs are well known examples of theories where additional
fermions and new neutral gauge bosons are simultaneously present.
For a general
breaking of E$_6$ (rank 6) to the SM (rank 4) it is possible to define
a whole class of $Z_1$ bosons corresponding to a linear combination of the two
additional Cartan generators.
I will parametrize this combination in terms of an
angle $\beta$. Fermions are assigned to the
fundamental {\underbar {\elevenbf 27}}
representation of the group which contains 12 additional states
for each generation, among which we have a vector doublet of new
leptons $(N\> E^-)_L^T$, $(E^+ \> N^c)_L^T$. Non-diagonal mass
terms with the standard $(\nu\> e^-)_L^T$ and $e^c_L$ leptons will
give rise respectively to ordinary-ordinary and ordinary-exotic
mixings, and in particular will induce LFV
{\elevenit L} and {\elevenit R} chiral
couplings between the first and second
\vfil\eject

\phantom{p}
\vskip6.5truecm
\midinsert
{
\vskip -.9truecm
\hskip-1.0truecm\includegraphics{Fig1.eps}
$$
\vbox{
\vskip -.5truecm
\hsize= 13.3truecm
{\tenrm \baselineskip=14pt
\noindent
Figure 1:
Limits on the $Z_1$ LFV parameter
$M_{Z_1}\cdot \left({\cal F}_L^{e\mu}\right)^{-1/2}$
from the limits on the $\mu$-$e$ conversion process\upcite{mue-exp},
for a general E$_6$  gauge boson, as a
function of $\sin\beta$.
The mixing term ${\cal F}_L^{e\mu}$ is in units
of $10^{-2}$, vertical units are TeV.
 Limits on the $Z_1$ mass for different values of
${\cal F}_L^{e\mu}$ can be easily read off the figure by
properly rescaling the vertical units.}
}
$$
}
\endinsert
\vskip -.6truecm
\noindent
generation, allowing for LFV processes as
$\mu\rightarrow eee$\upcite{lfc} and
$\mu$-$e$ conversion in nuclei\upcite{mue},
for which very stringent experimental limits exist\upccite{mue-exp}{sindrum}.
Due to the expected suppression of the ordinary-exotic mixings, I have
conservatively neglected
the LFV couplings in the {\elevenit R} sector, and assumed
that the only source of LFV interaction comes from the
ordinary-ordinary mixing in the {\elevenit L} sector\upcite{lfc}.
Since in this case the exchange of the $Z_0$ does not contribute
to the $\mu$-$e$ transition, the relevant parameter is actually
${\cal F}_L^{e\mu}/M_{Z_1}^2$.
Relying on an analogy with the CKM mixings, which  also
conserve weak-isospin and are all $> 10^{-3}$ and noting that the mixing
 between the first and second generation is particularly large,
I have assumed a `natural' range
$10^{-2}$--$10^{-3}$ for the LFV term
${\cal F}_L^{e\mu}$.
The corresponding limits on the $Z_1$ mass
obtained from the limit\upcite{mue-exp}
$B(\mu N\to e N) < 4\times 10^{-12}$
are depicted in Fig.$\, 1$ which is taken from
Ref. [\cite{mue}].
The limits are given as a function of the
parameter $\beta$ that defines the particular E$_6$ boson,
and are more stringent than the corresponding limits implyed
by $\mu \not\to eee$\upcite{lfc}.
It is apparent that these
limits are indeed very strong, however,
since they depend on a specific assumption for the numerical
value of the LFV coupling, they clearly cannot replace
the direct\upcite{zp-direct} bounds or other more
model independent indirect limits\upccite{fit6}{zp-new}.
\vglue 0.6cm
\adv{1}
\line{\elevenbf 4. Acknowledgements \hfil}
\vglue 0.4cm
\noindent
It is a pleasure to thank my friends and collaborators E. Roulet
and D. Tommasini, with whom large part of this work was done.
I would also like to thank D. London for helpful discussions.
\vglue 0.6cm
\line{\elevenbf 5. References \hfil}
\vglue 0.4cm

\biblitem{taumass}
BEPC collaboration, J. Z. Bai \ea, \prl{69} (1992) 3021.\par

\biblitem{taulife}
W. Trischuk, in ``{\elevenit Proc. of the Second Workshop on Tau Lepton
Physics}"
(The Ohio State Univ., Columbus) ed. K. K. Gan (1992) to be published. \par

\biblitem{taubranch}
R. Galik, in Ref. [\cite{taulife}].\par

\biblitem{UA1-B}
UA1 Collaboration, C. Albajar \ea, \plb{262} (1991) 163.\par

\biblitem{mirror}
See, for example, F. Wilczek and A. Zee, \prd{25} (1982) 553;\hbup
J. Bagger \ea, \npb{244} (1984) 247.  \par

\biblitem{lep-lfc}
L3 collaboration, B. Adeva \ea, \plb{271} (1991) 453; \hbup
OPAL Collaboration, M. Akrawy \ea, \plb{254} (1991) 293; \hbup
ALEPH Collaboration, D. Decamp \ea, \prep{216}(1992) 253; \hbup
DELPHI Collaboration, P. Abreu \ea, \plb{298} (1993) 247. \par

\biblitem{fit}
E. Nardi, E. Roulet and D. Tommasini,  \npb{386} (1992) 239. \par

\biblitem{ll1}
P. Langacker and D. London, \prd{38} (1988) 886.\par

\biblitem{ll2}
P. Langacker and D. London, \prd{38} (1988) 907.\par

\biblitem{london}
D. London, Report No. UdeM-LPN-TH131,
to be published in {\elevenit
Precision Tests of the Standard Model}, 1993, ed.\ P. Langacker (World
Scientific).\par

\biblitem{fit6}
E. Nardi, E. Roulet and D. Tommasini, \prd{46} (1992) 3040. \par

\biblitem{lfc}
E. Nardi, \prd{48} (1 August 1993). \par

\biblitem{mue}
J. Bernab\'eu, E. Nardi and D. Tommasini,
UM-TH-93-08 (May 1993).  \par

\biblitem{trip}
B.W. Lee, \prd{6} (1972) 1188; \hbup
J. Prentki and B. Zumino, \npb{47} (1972) 99; \par 

\biblitem{Silverman}
D. Silverman, \prd{45} (1992) 1800. \par

\biblitem{pdg92}
Particle Data Group, J.J. Hern\'andez \ea, \prd{45} Part. II (1992). \par

\biblitem{sindrum}
SINDRUM collaboration, U. Bellgardt \ea, \npb{299} (1988) 1. \par

\biblitem{mue-exp}
TRIUMF collaboration,
S. Ahmad \ea, \prd{38} (1988) 2102; \hbup
PSI collaboration, A. Badertscher \ea, J. Phys. {\elevenbf G17} (1991)
S47;\hbup
PSI collaboration, A. Van der Schaaf, Nucl. Phys.
{\elevenbf A546} (1992) 421C.\par

\biblitem{zp-direct}
CDF Collaboration, F. Abe \ea, \prl{67} (1991) 2609; \ib {\elevenbf 68},
(1992) 1463. \par

\biblitem{zp-new}
P. Langacker and M. Luo, \prd{45} (1992) 278; \hbup
J. Layssac, F.M. Renard and C. Verzegnassi, \zpc{53} (1992) 97; \hbup
M.C. Gonzalez Garc\'\i a and J.W.F. Valle; \plb{259} (1991) 365; \hbup
G. Altarelli \ea, \plb{263} (1991) 459; \hbup
F. del Aguila \ea, \npb{372} (1992) 3. \par

\insertbibliografia

\vfill\eject\bye